\documentclass[12pt]{article}
\usepackage{epsfig,pstricks}

\newcommand{\nc}{\newcommand}
\nc{\beq}{\begin{equation}}
\nc{\eeq}{\end{equation}}
\nc{\beqa}{\begin{eqnarray}}
\nc{\eeqa}{\end{eqnarray}}

\newwrite\ffile\global\newcount\figno \global\figno=1

\def\writedef#1{}

\input epsf
\def\figin{\epsfcheck\figin}\def\figins{\epsfcheck\figins}
\def\epsfcheck{\ifx\epsfbox\UnDeFiNeD
\message{(NO epsf.tex, FIGURES WILL BE IGNORED)}
\gdef\figin##1{\vskip2in}\gdef\figins##1{\hskip.5in}
\else\message{(FIGURES WILL BE INCLUDED)}%
\gdef\figin##1{##1}\gdef\figins##1{##1}\fi}

\def\figinsert{}
\def\ifig#1#2#3{\xdef#1{fig.~\the\figno}
\writedef{#1\leftbracket fig.\noexpand~\the\figno}%
\figinsert\figin{\centerline{#3}}\medskip\centerline{\vbox{\baselineskip12pt
\advance\hsize by -1truein\center\footnotesize{  Fig.~\the\figno.} #2}}
\bigskip\endinsert\global\advance\figno by1}
\def\endinsert{}

\begin{document}

\title{\large{\bf 
Color Superconductivity in High Density Quark Matter
}}

\author{
Stephen D.H.~Hsu\thanks{hsu@duende.uoregon.edu} \\
Department of Physics, \\
University of Oregon, Eugene OR 97403-5203 \\ \\   }

\date{March, 2000}

\maketitle

\begin{picture}(0,0)(0,0)


\end{picture}

\vspace{-24pt}

\begin{abstract}
We review recent progress on the phenomena of color superconductivity
in high density quark matter. We begin with a brief overview of the unique 
aspects of physics near a Fermi surface and the implications for
renormalization group (RG) techniques. We next discuss the qualitative 
differences between asymptotic densities, where the effective coupling 
constant can be made arbitrarily small, and intermediate densities where 
quark matter is still strongly coupled. It is in the latter regime where 
RG techniques are particularly useful, in that they yield a generic
description of possible behaviors without relying on an expansion in the
strong coupling constant. Finally, we discuss aspects of the QCD 
groundstate at asymptotic densities, which can be determined in a 
systematic weak coupling expansion. 
\end{abstract}

\bigskip

\begin{center}
{\it Contribution to the Proceedings of the TMU-Yale Symposium
on the Dynamics of Gauge Fields, December 13-15 1999, Tokyo, Japan.}
\end{center}

\newpage

\section{Introduction}

\begin{figure}[htb]
\center{
\leavevmode
\epsfbox{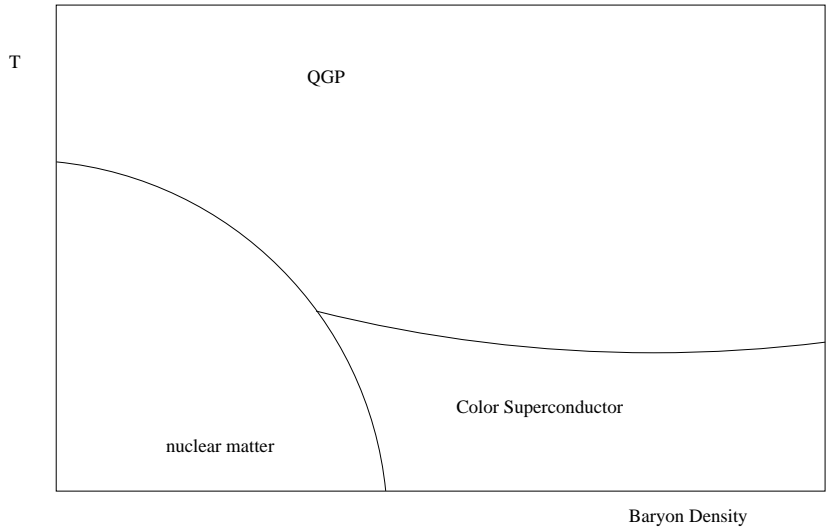}
\caption{The QCD Phase Diagram as a function of temperature and
density.} 
\label{phasediag}
}
\end{figure}

In this article I review recent progress on color superconductivity
in high density QCD. It is remarkable how much more we know about
QCD at high density compared to just two years ago, in 1998. The
ultimate fate of baryonic matter at high density is a fundamental
property of QCD, with implications for the astrophysics of neutron
stars, as well as heavy ion collisions. More generally, we would
like someday to understand the QCD phase diagram 
(figure (\ref{phasediag}))
in all its complexity
as a function of temperature, chemical potential and quark masses.

While lattice studies have been fundamental in determining the
behavior of QCD at high temperature, technical difficulties
arise in the application of Monte Carlo techniques once a chemical potential
is introduced (essentially, the measure of integration is no longer
positive definite, due to complex eigenvalues of the Dirac operator). 
The progress I describe does not rely on brute
force techniques, but rather on physical insights associated with
the presence of a Fermi surface. The special
properties of physics near a Fermi surface are discussed in section 2. 
Remarkably, rigorous statements can be made in the limit of 
infinite density.

\epsfysize=7 cm

There are two qualitatively different regimes to be addressed.
The regime most likely to be realized in nature (i.e. in the cores of
neutron stars or in heavy ion collisions) 
is that of intermediate density, where the average distance
between quarks is still of order a Fermi and the gluons exchanged
are rather soft. This regime is strongly coupled, and while we can
make qualitative statements about the nature of the ground state using
renormalization group techniques, we cannot perform quantitative 
calculations. The intermediate density case is discussed in section 2.

At extremely high densities the effective coupling 
$\alpha_s$ is small, leading to a weakly coupled liquid
of quarks. The only low energy excitations in this liquid are quasi-particles
and quasi-holes representing fluctuations near the Fermi surface.
There are rather general arguments (see below) which suggest that 
any attractive interaction, no matter how small, can lead to Cooper
pairing in the presence of a Fermi surface. In the case of QCD, this
attractive interaction is provided by gluon exchange in 
the $\bar{3}$ channel. Thus, we expect to find condensation of Cooper 
pairs of quarks in the high density limit. 

Actually, even in the high density limit, there are some subtleties 
which make the problem less than straightforward. Most importantly,
although the effective coupling is weak, the exchange of magnetic
gluons leads to a long range interaction due to the absence of a
magnetic screening mass. Quarks carry electric, rather than
magnetic, color charge, and hence are better at screening the timelike
component ($A_0$) of the gluon field than the spacelike ($A_i$). It
is only non-perturbative effects which can screen color-magnetic 
fluctuations, and these are nearly absent at weak coupling. An 
understanding of dynamic screening due to Landau damping is necessary
to control the high density calculations. These issues are discussed
in section 4.

The order parameter for color superconductivity is
\begin{equation}
\langle \psi^T C \, \Gamma \, \psi \rangle~~~.
\end{equation}
where C is the charge conjugation operator and
$\Gamma$ is a matrix in color, flavor and Dirac space.
As we will see below, determining the precise form of 
$\Gamma$ requires some work; it is particularly complicated
in the case of three flavors. However, as we will discuss in
section 5, in the weak coupling limit the true groundstate
can be determined in a controlled approximation.
Here we will simply note that
single gluon exchange between two quark quasiparticles can
be decomposed into an attractive $\bar{3}$ channel and a
repulsive $6$ channel. Thus, at the most naive level we
expect an anti-triplet condensate, which breaks 
$\rm SU(3) \rightarrow SU(2)$. The main effect of this 
condensate is that it leads to the Higgs phenomena for at
least some subset of the gluons, or equivalently to the
Meissner effect and screening of some subset of the color
magnetic fields. The phenomenological implications of 
color superconductivity are still not well understood and
will be the subject of investigation for some time to come.

Let me close this introduction with a historical note. The
idea that quark matter might be a color superconductor is
quite old \cite{hsuFrau,hsuBarr,hsuBL}. The original insight
was based on the existence of the attractive $\bar{3}$ channel
and an analogy with ordinary superconductors. Recent interest 
in the problem was rekindled by the work of two groups that 
considered diquark condensation due to instanton-mediated 
interactions \cite{hsuI}, predicting gaps as large as
$\sim 100$ MeV. These calculations, while uncontrolled, are
quite suggestive, and led to the recent progress on the subject.
It is often claimed that early investigations predicted tiny gaps, 
at most of order a few MeV. While this may have been the consensus 
among the few theorists who had actually worked on the problem, it is 
actually an unfair characterization of Bailin and Love's results
\cite{hsuBL}. A value of the strong coupling large enough to 
justify the instanton liquid picture of \cite{hsuI} also yields a 
large gap when substituted in Bailin and Love's results. After all, 
instantons are suppressed by an exponential factor
$\exp( - 2 \pi / \alpha_s )$. Bailin and Love merely suffered from the
good taste not to extrapolate their results to large values of $\alpha_s$!

\section{Physics Near a Fermi Surface and the Renormalization Group}

Important simplifications arise in the study of cold, dense matter due
to the existence of a Fermi surface, which in relativistic systems is 
likely to be rotationally invariant. The energy of a low-energy 
excitation (a quasi-particle or -hole) is then independent of the
orientation of its momentum $\vec{p}$, and only depends on 
$p - \mu$, where $p = | \vec{p}|$ and $\mu$ is the chemical potential
or Fermi energy. (Here, for simplicity, we will always work with
massless quarks.) This leads to a kind of dimensional reduction, so
that physics near a Fermi surface is effectively 1+1 dimensional.
In particular, arbitrarily weak interactions can lead to non-perturbative
phenomena like pair formation.

The renormalization group approach \cite{hsuRGold} 
is particularly useful here --
we integrate out the modes far from the Fermi surface, leaving only
the low-energy quasi-particle and -hole states that are involved 
in the interesting physics. These excitations might in principle be
related to the original quarks in a complicated way, but on
quite general grounds must be described by an effective action of 
the form
\begin{equation}
\label{Seff}
 S_{eff} = \int dt\, d^3p\; \psi^\dagger \left( i\partial_t
 - (\epsilon(p)-\epsilon_F) \right)\psi + S_{int},
\end{equation}
where $S_{int}$ contains higher dimensional, local quasi-particle
operators. 
Strictly speaking, this form of the effective action is only
valid for models in which the original interactions were
local (short ranged). While appropriate for QCD at intermediate
densities \cite{hsuRGEHS,hsuRGSW}, 
where non-perturbative effects are expected to 
generate screening of magnetic gluons, it must be modified 
at weak coupling where magnetic fluctuations are long ranged
\cite{hsuSon,hsuHS}. However, it is the only
technique I know of from which we can obtain robust information
about the strongly coupled region of the phase diagram. 
Below I review the results of this analysis, and defer a discussion
of the weak coupling phase until the following section.

It can be shown \cite{hsuRGEHS,hsuRGSW} using simple
classical scaling arguments that all interactions are irrelevant
except for the Cooper pairing interaction (scattering of
quasi-particles at opposide sides of the Fermi surface: 
$\vec{p}_1 \simeq - \vec{p}_2$) and stricly colinear scattering:
$\vec{p}_1 \simeq \vec{p}_2$, which can lead to the 
Overhauser effect (chiral waves) at large-$N_c$ \cite{hsuOver}. 
Both of these interactions are classically marginal, so quantum
corrections determine their evolution. Here we restrict
ourselves to local Cooper pairing operators which are 
invariant under the full
$SU(3)_L\times SU(3)_R\times U(1)_A$ chiral symmetry:
\begin{eqnarray}
\label{ops}
O^0_{LL} &=& (\bar\psi_L\gamma_0\psi_L)^2, \hspace{1cm}
O^0_{LR} \;=\; (\bar\psi_L\gamma_0\psi_L)(\bar\psi_R\gamma_0\psi_R) \\
O^i_{LL} &=& (\bar\psi_L\gamma_i\psi_L)^2,  \hspace{1cm}
O^i_{LR} \;=\; (\bar\psi_L\vec\gamma\psi_L)
       (\bar\psi_R\vec\gamma\psi_R). \nonumber
\end{eqnarray}
These come in both color symmetric ($\bar{3}$) and antisymmetric
(sextet) combinations.
More general operators with different flavor or Dirac structures 
can be reduced to linear combinations of the basic ones 
(\ref{ops}), using parity and Fierz rearrangements. 
This analysis can be extended to operators (such as those
induced by instantons) that break the anomalous $U(1)_A$ symmetry 
\cite{hsuRGEHS,hsuRGSW}, yielding a very robust characterization
of QCD even at intermediate densities and strong coupling. 
We will not discuss the details of this more general analysis here,
but the results (given reasonable assumptions about the signs and 
magnitudes of the interactions) are qualitatively similar 

The RG evolution of the operators in (\ref{ops}) is determined
by quark-quark scattering near the Fermi surface. 
A bubble graph with four-quark vertices $\Gamma_1$ and $\Gamma_2$ 
and external quark lines satisfying the Cooper pairing kinematics
yields 
\begin{equation}
\label{bubble}
G_1 G_2 I\; (\Gamma_1)_{i'i}(\Gamma_1)_{k'k}
  \left[ -(\gamma_0)_{ij}(\gamma_0)_{kl}+\frac{1}{3}
          (\vec\gamma)_{ij}(\vec\gamma)_{kl}\right]
 (\Gamma_2)_{jj'}(\Gamma_2)_{ll'}
\end{equation}
Here $I=\frac{i}{8\pi^2}\mu^2\log(\Lambda_{IR}/\Lambda_{UV})$, 
where $[\Lambda_{IR},\Lambda_{UV}]$ are the upper and lower
cutoffs of the momentum shell integrated out. 
We define the density of states on the Fermi surface to be  
$N=\mu^2/(2\pi^2)$ (in weak coupling) and 
$t \equiv \log(\Lambda_{IR}/\Lambda_{UV})$. The RG flow 
does not mix $LL$ and $LR$ operators, nor different color channels. 
We obtain the following RG equations
\begin{eqnarray}
\frac{d(G^{LL}_0+G^{LL}_i)}{dt} &=& -\frac{N}{3}
   (G^{LL}_0+G^{LL}_i)^2, \label{E1} \\
\frac{d(G^{LL}_0-3G^{LL}_i)}{dt} &=& -N
   (G^{LL}_0-3G^{LL}_i)^2, \label{E2} \\
\frac{d(G^{LR}_0+3G^{LR}_i)}{dt} &=& 0,\\
\frac{d(G^{LR}_0-G^{LR}_i)}{dt} &=& -\frac{2N}{3}
   (G^{LR}_0-G^{LR}_i)^2.
\end{eqnarray}
The linear combination 
$G_* = G^{LL}_0+G^{LL}_i$ reaches its Landau pole first, 
governed by the equation
\begin{equation}
 G_*(t) = \frac{1}{1+(N/3)G_*(0)t}~~~.
\end{equation}
In general, interactions which are attractive at the matching scale, 
$G_*(0)>0$, will grow during the evolution, and reach a Landau pole 
at the scale $t_* = - 3/(N G_* (0))$. The corresponding energy scale is 
\begin{equation} 
\Lambda_{IR} = \Lambda_{UV} \exp\left(-\frac{3}{N G_* (0)}\right),
\end{equation}
which agrees with the usual BCS result. Repulsive interactions,
corresponding to negative initial values of the coupling, become
weaker near the Fermi surface.

\epsfysize=7 cm
\begin{figure}[htb]
\center{
\leavevmode
\epsfbox{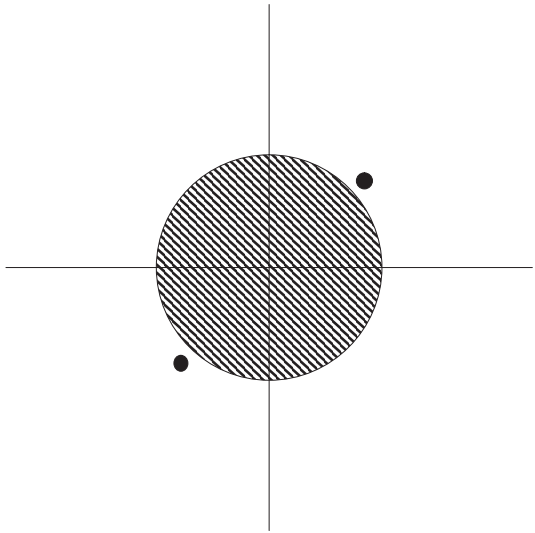}
\caption{Excitations on opposite sides of the Fermi surface.}
\label{FS}
}
\end{figure}

\section{High Density Limit}

In the high density limit the typical momentum transfer between
quarks is large, and therefore the effective coupling is small.
The properties of this phase can be deduced in a systematic,
weak coupling expansion. However, there is one technical
problem that must be solved having to do with soft magnetic gluons.
In the renormalization of the Cooper pairing interaction there is
a region of phase space where the incoming quarks are only slightly
deflected by the gluon exchange. This leads to an IR divergence unless
arbitrarily soft gluons are screened in some way. While gluons
acquire a perturbative electric mass at high density, it can be shown
that to all orders in perturbation theory no magnetic mass is 
generated \cite{hsuLB}. The only hope is that Landau damping -- a 
form of dynamic screening affecting the spacelike gluons -- 
is enough to control this IR problem. That this is so was first pointed
out by Son \cite{hsuSon}, who went on to deduce the following behavior
for the diquark gap at weak coupling
\begin{equation}
\label{LP}
\Delta \sim \mu g^{-5} \exp \Bigl( - {3 \pi^2 \over \sqrt{2} g} \Bigr) ~~~.
\end{equation}
This result has since been confirmed by RG methods \cite{hsuHS} as
well as Schwinger-Dyson techniques \cite{hsuweakgap}. In this section
I discuss some of the details of these calculations, concentrating on
the the RG approach.

The magnetic gluon propagator, including vacuum
polarization effects from virtual quarks, \cite{hsuLB} has the 
form ($q_0 << q$)
\begin{equation}
\label{LDD}
D^T_{\mu \nu} (q_0, q) 
~=~ { P^T_{\mu \nu} \over q^2 + i\frac{\pi}{2} m_D^2 \frac{|q_0|}{q} }~~~.
\end{equation}
Strictly speaking $D^T_{\mu \nu}$ is gauge dependent, but in our
leading order calculations the propagator always appears contracted
with gamma matrices next to nearly on-shell external quark lines. Thus the
gauge dependent parts are higher order in g due to the equations of motion.
The effect of Landau damping is to cut off the small-q divergence in
(\ref{LDD}) at $q \sim q_0^{1/3} m_D^{2/3}$, where
$m_D^2 = N_f {g^2 \mu^2 \over 2 \pi^2}$ is the Debye screening mass.
A common feature of both the Schwinger-Dyson and RG calculations in
the weak coupling region is loop integrals dominated by energy transfers
of order $q_0 \sim \Delta$, and hence momentum
transfers of order $q_* \sim \Delta^{1/3} m_D^{2/3}$. 
$q_*$ can be made as large as desired by going to high density.

The main technical problem in the RG approach is long range magnetic
interactions, or equivalently the presence of soft gluons. At no
point can the theory be completely described by quarks with purely
local interactions as in (\ref{Seff}). The effective Lagrangian contains 
both quark and gluon excitations (with energies below the cutoff $\Lambda$) 
and local interactions resulting from integration of higher energy shells.
This modifies the form of the RG equations obtained \cite{hsuHS},
so that equations (\ref{E1}) and (\ref{E2}) become
\begin{eqnarray}
\frac{d(G^{LL}_0+G^{LL}_i)}{dt} &=& -\frac{N}{3}
   (G^{LL}_0+G^{LL}_i)^2 ~-~ {g^2 \over 9 \mu^2} ~~~ , \label{E1mod} \\
\frac{d(G^{LL}_0-3G^{LL}_i)}{dt} &=& -N
   (G^{LL}_0-3G^{LL}_i)^2 ~+~ {g^2 \over 27 \mu^2}~~~  . \label{E2mod} \\
\nonumber
\end{eqnarray}
The solution of these RG equations leads to a Landau pole in the
dominant $\bar{3}$, LL and RR channels given by (\ref{LP}). It is worth
commenting on the angular momentum of the condensate. The RG equations can
be derived for general values of angular momentum l. Naively interpreted, the
results suggest that condensates might occur in higher l channels, leading
to the breaking of O(3) rotational invariance. A more detailed gap equation
analysis \cite{hsuHS} shows that this is not the case: a large s-wave gap
suppresses the formation of p-wave and higher l gaps. The literature is
somewhat confused on this important issue. The papers in 
\cite{hsurotation} address the issue of rotational invariance and do not
agree on the size of higher l gaps. However, neither paper addresses the
interplay of s-wave and higher l gaps, which is studied in \cite{hsuHS}.

\section{The QCD Groundstate at High Density}

In this section I describe the vacuum
energy analysis necessary to determine the groundstate of QCD
at high density \cite{hsuEHHS}. 
Neither the RG nor Schwinger-Dyson analyses are
sufficient to specify the actual groundstate. Strictly speaking,
the former only reveals the energy scale and quantum numbers of 
the pairing instability, while the latter only identifies extrema of 
the vacuum energy. As we shall see, there are additional subtleties
which can only be resolved by consideration of energetics.

\begin{figure}[htb]
\center{
\leavevmode
\epsfbox{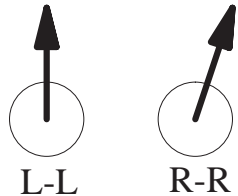}
\caption{Color disorientation of LL and RR condensates.}
\label{diso2}
}
\end{figure}

First, let us consider the case of 2 massless flavors.
Because the condensate occurs between pairs of either 
left (LL) or right (RR) handed
quarks in the J=L=S=0 channel \cite{hsuHS}, and the $\bar{3}$ color
channel is antisymmetric, the quarks must pair in the 
isospin singlet (ud - du) flavor
channel. However, even in this case there is a subtlety, as
the relative color orientations of the LL and RR condensates are 
not determined by the usual leading order analysis. A misalignment
of these condensates violates parity,
and further breaks the gauge group beyond 
$\rm SU(3)_c \rightarrow SU(2)_c$.
An analysis of the Meissner effect is 
necessary to determine the relative orientation \cite{hsuEHHS},
and the effect is higher order in g.
There are thus a number of unstable configurations of only slightly
higher energy with
different color-flavor orientations (and hence different symmetry
breaking patterns), leading to the possibility of
disorienting the diquark condensate (see figure (\ref{diso2})).

\epsfysize=2.5 cm

\epsfysize=2.5 cm
\begin{figure}[htb]
\center{
\leavevmode
\epsfbox{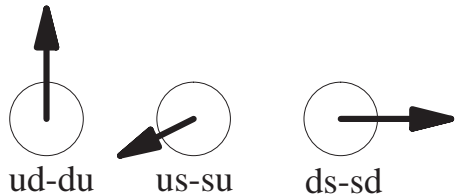}
\caption{CFL condensation: each flavor channel orients chooses
an orthogonal color orientation.}
\label{diso3}
}
\end{figure}

The generalization to three flavors is far from straightforward. 
Again, one can show that
the condensate must occur in the J=L=S=0 and color $\bar{3}$ channel.
(The sextet condensate is suppressed
in the weak coupling limit \cite{hsuEHHS} 
and I do not discuss it here.)
The Pauli principle
then requires that the flavor structure again be antisymmetric 
$\sim ~(q_i q_j - q_j q_i)$, for quarks of flavor $i,j$. Thus, 
one can have combinations
of condensates which are in the $\bar{3}$ of both color 
and flavor $\rm SU(3)_L$
or $\rm SU(3)_R$. Due to the chirality preserving nature of 
perturbative gluon exchange, 
there is no mixing of LL and RR condensates, which form independently. 
One can immediately
see that there are a number of possibilities. For example, the 
condensates for the three flavors and both chiralities might all align
in color space, leading to an $\rm SU(3)_c \rightarrow SU(2)_c$ 
breaking pattern. A more
complicated condensate has been proposed \cite{hsuARW2,hsuSW2} called 
Color Flavor Locking (CFL),
in which the $\bar{3}$ color orientations are ``locked'' to 
the $\bar{3}$ flavor orientation. In figure (\ref{diso3}) we give
a simple picture of CFL condensation. 
 
To determine the nature of the energy 
surface governing 
the various color-flavor orientations of the condensate, we can
begin by characterizing
the color-flavor configuration space of condensates. We 
consider the ans\"atz
\begin{equation}
\label{ansatz}
\Delta^{ab}_{ij}~{}^{L,R} ~=~ A^c_k ~{}^{L,R}~ 
\epsilon^{abc} \epsilon_{ijk} ~~~,
\end{equation}
where a,b are color and i,j flavor indices. L and R denote 
pairing between pairs of
left and right handed quarks, respectively. Under color and flavor A
transforms as
\begin{equation}
\label{Atrans}
A^{L} \rightarrow   U_c A^L V^L~~~,
\end{equation}
where $U_c$ is an element of $\rm SU(3)_c$ and $V^{L}$ of 
$\rm SU(3)_{L}$. A similar equation holds for $A^R$. 
It is always possible to diagonalize $A^L$ by appropriate 
choice of $U_c$ and $V^L$:
\begin{equation}
\label{ansatz1}
A^L  ~=~ 
\left(\begin{array}{ccc}
a & 0 & 0 \\
0 & b & 0 \\
0 & 0 & c 
\end{array} \right)~~~.
\end{equation}
Generically, there does not exist a $V^R$ which diagonalizes
$A^R$ in this basis. In the CFL case, where the diagonalized 
$A^L$ is proportional to the identity, $a=b=c$, it is easy to show that 
one can choose $V^R$ such that $A^R = \pm A^L$. These two configurations
are related by a $U(1)_A$ rotation (see section 3). Hence, they are degenerate
in the high density limit where gluon exchange dominates. Instanton effects,
important at intermediate density, favor $A^R = A^L$.
Note that parity, if unbroken, requires 
$A^L = A^R$, and hence implies simultaneous diagonalizability.

In \cite{hsuEHHS} we considered the potential vacua parametrized by a,b,c.
First, we use the Dyson-Schwinger (gap) equation to determine which of these 
configurations are energy extrema. Next, we computed the energies of the
extrema to determine the true groundstate. A similar analysis has been
carried out by Sch\"afer and Wilczek \cite{hsuSW2} in the 
approximation where gluon
interactions are replaced by local four fermion
interactions. They concluded that the CFL vacuum had the lowest energy.
In our analysis, which I summarize below, 
we included the gluons in the analysis, introducing
long range color-magnetic fluctuations (controlled by Landau damping) 
and Meissner screening into the gap equation and vacuum energy calculations.

\begin{figure}[htb]
\center{
\leavevmode
\epsfbox{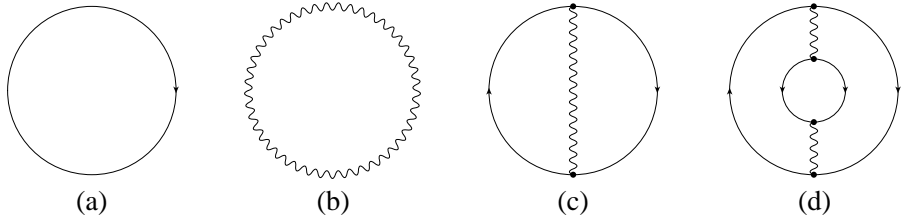}
\caption{Vacuum energy diagrams.}
\label{fig1}
}
\end{figure}

\begin{figure}[htb]
\center{
\leavevmode
\epsfbox{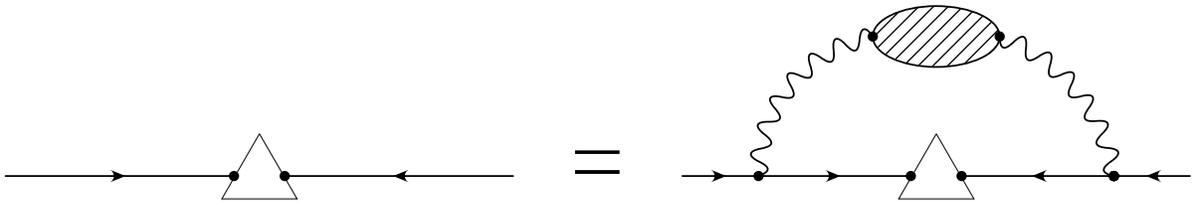}
\caption{Schwinger-Dyson equations.}
\label{fig2}
}
\end{figure}

At asymptoticaly high densities (weak coupling) the diagrams (a)-(c) in 
figure (\ref{fig1}) give the leading approximation to the effective action. 
Note that in these diagrams the quark propagators include the diquark
condensate (see (\ref{SI}) below), and the gluon propagators include
Landau damping, but {\it not} the Meissner effect. The latter arises
from the condensate-dependence of quark loops in diagrams (c) and (d). 
The resulting gap equation (figure (\ref{fig2})), with condensate shown
explicitly at lowest order in $\Delta$) is given by
\begin{equation}
\label{gapeqn1}
S^{-1}(q) - S^{-1}_0 (q) ~=~ i g^2 \int {d^4k \over (2 \pi)^4} 
  ~\Gamma^A_\mu  
~S(k)~ \Gamma^B_\nu  ~D^{\mu \nu}_{AB} (k-q)~,
\end{equation}
where 
\begin{equation}
\label{G}
\Gamma^A_\mu~=~ \left( \begin{array}{cc}
\gamma_\mu T^A  & 0 \\
0 & C (\gamma_\mu T^A)^T C^{-1}
\end{array} \right)~~~.
\end{equation}
$~D^{\mu \nu}_{AB}$ is the gluon propagator, including the
effects of Landau damping and Debye screening (we assume
Feynman gauge throughout).

We will restrict the color group structure in the gap equation
to the attractive anti-symmetric $\bar{3}$ channel,
which projects out the anti-symmetric part of $S(k)$ in color space in
the gap equation.
Here $S$ is the fermion propagator for the spinor $(\psi^i_a, \psi^{i C}_a)$ 
with $i$ a flavor
index and $a$ a color index.

\epsfysize=3 cm

\epsfysize=3 cm

For the three flavor case
$S$ can be written explicitly as an 
$18 \times 18$ matrix in color flavor space. 
The inverse propagator may be written
\begin{equation}
\label{SI}
S^{-1}(q) = \left( \begin{array}{cc}
q\!\!\!/ + \mu\!\!\!/  & \gamma_0 \Delta^\dagger \gamma_0 \\
\Delta  &  q\!\!\!/ - \mu\!\!\!/ 
\end{array} \right)
\end{equation}
where $\mu\!\!\!/ = \mu \gamma_0$.
$\Delta$ is a $9 \times 9$ matrix which for the ans\"atz (\ref{ansatz1}) 
takes the form
\begin{equation} \label{cond}
\Delta = 
\left( \begin{array}{ccccccccc}
   0 & 0 & 0 & 0 & c & 0 & 0 & 0 & b \\
   0 & 0 & 0 & -c & 0 & 0 & 0 & 0 & 0 \\
   0 & 0 & 0 & 0 & 0 & 0 & -b & 0 & 0 \\
   0 & -c & 0 & 0 & 0 & 0 & 0 & 0 & 0 \\
   c & 0 & 0 & 0 & 0 & 0 & 0 & 0 & a \\
   0 & 0 & 0 & 0 & 0 & 0 & 0 & -a & 0 \\
   0 & 0 & -b & 0 & 0 & 0 & 0 & 0 & 0 \\
   0 & 0 & 0 & 0 & 0 & -a & 0 & 0 & 0 \\
   b & 0 & 0 & 0 & a & 0 & 0 & 0 & 0 
 \end{array} \right)
\end{equation}
Because we are dealing with a diquark condensate
the non-trivial part of the gap equation involves the 
lower left $9 \times 9$ block. We will refer to this sub-block of the 
propagator $S$ as $S_{21}$. 

For a particular ans\"atz $\Delta$ to be a solution to the gap equation we 
require that the color antisymmetric part of 
$T^A~ S_{21}(k) ~T^A$ (corresponding to the $\bar{3}$ channel) 
be proportional in color-flavor space to the off-diagonal 21
submatrix of $S^{-1}(q) - S^{-1}_0(q)$, or $\Delta (q)$, 
which appears on the LHS of the gap equation (see \cite{hsuEHHS}
for more discussion of this point). 

The propagator may be found by inverting the
sparse matrix in (\ref{SI}) using Mathematica. 
Only three ans\"atze satisfy our condition:
$a=b=c$; $a=b, c=0$; $b=c=0$. We refer to these solutions as
(111) (color-flavor locking), (110) ($3 \rightarrow 0$ breaking) 
and (100) ($3 \rightarrow 2$ breaking) respectively.
For these ans\"atze the color antisymmetric part of 
$T^A S_{21}(k) T^A$ has the form of a constant
multiplying the matrix form (\ref{cond}) with $a,b,c$ set
to 0 or 1 as is appropriate for the ans\"atz. 
After contour integration over l, we find the following gap kernels
\begin{eqnarray}
(111) ~&:&~ \frac{2}{3} {\Delta \over \sqrt{k_0^2 + \Delta^2} } +  \frac{1}{3}
{\Delta \over \sqrt{k_0^2 + 4 \Delta^2} } \nonumber \\
(110) ~&:&~ { \Delta \over 2 \sqrt{k_0^2 + \Delta^2} } +
{ \Delta \over 2 \sqrt{k_0^2 + 2 \Delta^2} } \nonumber \\
(100) ~&:&~ { \Delta \over \sqrt{k_0^2 + \Delta^2} }
\label{kernels}
\end{eqnarray}
These kernels are to be substituted in the following gap equation,
which we obtain under the approximation $q_0 << |\vec{q}|$. (We
also neglected the anti-particle contributions, which are suppressed
by powers of $1/ \mu$.)

\newpage

\begin {eqnarray}
\label{finalgap}
\Delta(p_0) &=& \frac{g^2}{12\pi^2} \int dq_0\int d\cos\theta\,
 \left(\frac{\frac{3}{2}-\frac{1}{2}\cos\theta}
            {1-\cos\theta+(G+(p_0-q_0)^2)/(2\mu^2)}\right. \\
 & & \hspace{3cm}\left.    +\frac{\frac{1}{2}+\frac{1}{2}\cos\theta}
            {1-\cos\theta+(F+(p_0-q_0)^2)/(2\mu^2)} \right)
K(q_0), \nonumber
\end {eqnarray}
where F and G represent the medium effects on the electric and 
magnetic gluons, 
and $K(q_0)$ is one of the gap kernels from (\ref{kernels}).
The Meissner effect makes an additional contribution to G beyond
that of Landau damping. In \cite{hsuEHHS} we evaluated 
the gluon vacuum polarization 
$P_{\mu \nu} (q_0,q)$ in the presence of a diquark condensate.
(A more detailed computation
of the Meissner effect is given by Rischke \cite{hsuDRM}, with
similar results.)
The additional Meissner screening is given by
$
\delta G ~\equiv~ {1\over 2} {\cal P}^T_{ij} P_{ij}~,
$
where
${\cal P}^T_{ij} = \Big(\delta_{ij} - \hat{q}_i \hat{q}_j\Big)$
is the transverse projection operator. At low momenta,
$q_0,q \sim \Delta$, $\delta G (q_0 ,q)$ 
is of order
the Debye mass $m_D \sim g \mu$, while at larger energy or
momenta the effect is suppressed by a power of ${\Delta \over q_0}$
or ${\Delta \over q}$.
We limited ourselves to an estimate of the
size of the Meissner effect on the gap solutions. To this end,
we used the following approximation for $\delta G$:
\begin{equation}
\delta G(q_0,q) \simeq m_D^2  {\Delta_0 \over  \sqrt{ q^2 + q_0^2 +
\Delta_0^2 } } ~~~ ,
\end{equation}
where $\Delta_0$ is the maximum value of the function $\Delta(k_0,k)$.
Note we did not introduce any color structure in $\delta G$; all
gluons experience the same magnetic screening. While this is a crude
approximation, it gives the rough size of the Meissner effect
on $\Delta$.

\begin{figure}[htb]
\center{
\leavevmode
\epsfbox{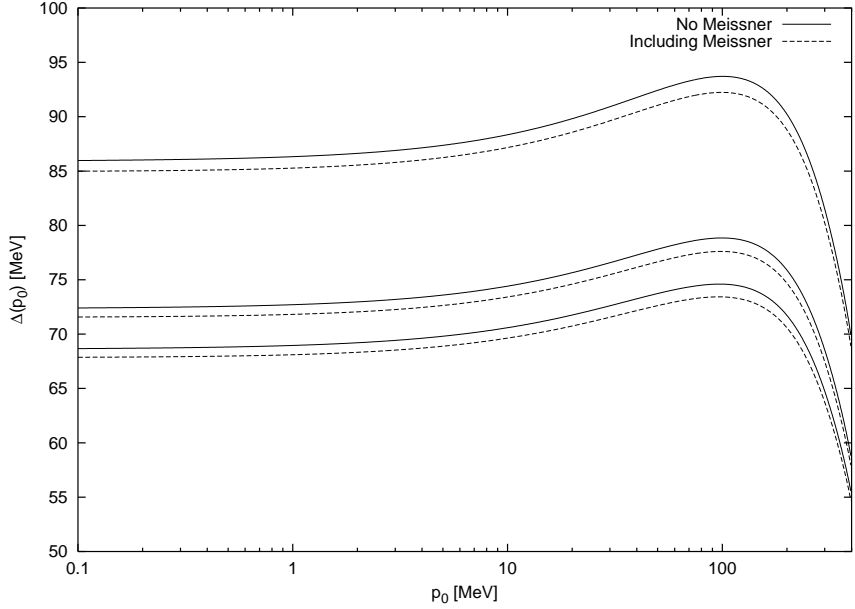}
\caption{Gap solutions for $\mu = 400$ MeV, with and without
Meissner effect.}
\label{fig5}
}
\end{figure}

We solved the gap equations for all three gap kernels using this
form of the Meissner effect, and 
the results are shown in figure (\ref{fig5}) for the case of
$\mu = 400$ MeV.  The effect is to decrease the size
of the condensate but it is a small perturbation on the solutions obtained 
without the Meissner effect.

\epsfysize=9 cm

To determine which of the above gaps is the true minimum energy state we 
must calculate the vacuum energy, which receives contributions from 
vacuum to vacuum loops of both quarks and gluons (figure 1). 
We start with the CJT effective potential \cite{hsuCJT}, which upon
extremization wrt appropriate propagators and 
vertices leads to the Schwinger-Dyson equations. The fermion equation
is the gap equation given above, while the gluon equation reproduces 
Landau damping. We wish to compare energies corresponding to our 
three solutions to 
determine which one is the true vacuum (the difference in
energies $V$ will be gauge invariant, whereas actual values are not). 
It is easy to show that the value of the effective potential 
evaluated on the gap solution is given by:
\begin{equation}
\label{cjtg}
V ~=~ -i \int {d^4p \over (2 \pi)^4} {\rm tr ln} S(p)/S_0(p)~.
\end{equation}
Diagramatically, this is equivalent to the graph of figure 1(a) when
evaluated on the gap solution.

The fermion
loops are most easily calculated by going to a basis where $S_0 S^{-1}$ 
is diagonal in color-flavor space. Note that the gap matrix $\Delta$
has non-trivial Dirac structure that must be accounted for \cite{hsuPRscalar}:
$\Delta = \Delta_1 \gamma_5 P_+ ~+~ \Delta_2 \gamma_5 P_-~$,
where $P_{\pm}$ are particle and anti-particle projectors. Our analysis
has been restricted to the particle gap function $\Delta_1$. The anti-particle
gap function $\Delta_2$ has its support near $k_0 \sim 2 \mu$, 
and its contribution to the vacuum energy is suppressed.
There are 18 eigenvalues, which occur in 9 pairs. The 
product of each pair is of the form 
\begin{equation}
 - \left( 1  ~+~ a { \Delta^2 ( k_o, k ) \over 
k_0^2 ~+~ (|\vec{k}| - \mu )^2 }  \right) 
~~~,
\end{equation}
where a is an integer. For our three cases we obtain the following
sets of eigenvalues:
\begin{eqnarray}
(111) ~~ &\rightarrow& ~~  8 \times \{ a=1 \} ~,~ 
1 \times \{a = 4 \} \nonumber \\ 
(110) ~~ &\rightarrow& ~~  4 \times \{ a=1 \} ~,~ 2 \times  \{ a = 2  \} 
\nonumber \\
(100) ~~ &\rightarrow& ~~  4 \times \{ a=1 \}
\end{eqnarray}

The binding energy is of order
\begin{eqnarray}
\label{Equark}
E_q & \sim &  - \int d^3k  dk_0 ~
\ln \left[ 1  + a { \Delta^2 (k_0,k) \over k_0^2 + (k - \mu)^2 } \right] \\
& \sim & - a ~\mu^2 \Delta_0^2 ~~, 
\end{eqnarray}
where $\Delta_0$ is the maximum value of the gap function $\Delta (k_0,k)$,
which has rather broad support in both energy and momentum space away from
the Fermi surface. A more precise answer
than (\ref{Equark}) requires numerical evaluation, but it is clear that the
result scales with a and has only a weak (logarithmic) dependence on the
variations in the shape of $\Delta (k_0,k)$. Substituting our numerical
results for the gaps in the three cases, it is easy to establish that
\begin{equation}
E(111) ~<~ E(110) ~<~ E(100)~~~.
\end{equation}

We find that the CFL vacua remains the lowest energy state,
at least at asymptotically high densities where the  calculation is reliable.
The Meissner effect is a small correction to the
vacuum energy at asymptotic densities. 
Configurations which satisfy the gap equations but are not the global
minimum of energy are presumably saddlepoints, since they are continuously
connected to the CFL vacuum via color and flavor rotations.

\section{Conclusions}

In this contribution I have tried to summarize some important
progress of the last two years on the theory of cold, dense quark
matter. Due to space limitations, I was not able to discuss a number
of important issues, such as the low energy effective Lagrangian,
continuity of hadronic and quark phases and more phenomenological
studies. I list some of the important papers in \cite{hsumore}.

I thank my collaborators N. Evans, J. Hormuzdiar and M. Schwetz 
and my colleagues D. Hong, R. Pisarski, K. Rajagopal, M. Rho,
D. Rischke and T. Sch\"afer for contributing to my understanding of
this subject. My research was supported under DOE contract DE-FG06-85ER40224
and by a JSPS visiting fellowship.

\end{document}